\documentclass[preprint,number]{elsarticle}

\usepackage{pgf}
\usepackage{tikz}
\usetikzlibrary{arrows,automata}
\usepackage[utf8]{inputenc}
\usepackage{amssymb}
\usepackage{amsthm}
\usepackage{amsmath}
\usepackage{mathtools}

\usepackage{hyperref}

\usepackage{mathpazo}
\newtheorem{thm}{Theorem}[section]
\newtheorem{lem}[thm]{Lemma}
\newtheorem{cor}[thm]{Corollary}
\newtheorem{prop}[thm]{Proposition}
\newtheorem{exam}[thm]{Example}

\newtheorem{que}[thm]{Question}
\newtheorem{defn}[thm]{Definition}

\newcommand{\N}{\mathbb{N}}

\newcommand{\cb}[1]{%
\left\{{#1}\right\}
}
\newcommand{\rb}[1]{%
\left({#1}\right)
}
\newcommand{\abs}[1]{%
\left|#1\right|
}

\newcommand{\fsg}[1]{%
{#1}^+
}
\newcommand{\fm}[1]{%
{#1}^*
}
\newcommand{\fg}[1]{%
\operatorname{F}\left({#1}\right)
}
\newcommand{\pows}[1]{%
\widehat{#1}
} 
\newcommand{\unitadd}[1]{%
{#1}^{e}
}
\newcommand{\zeroadd}[1]{%
{#1}^{z}
}
\newlength{\arrow}
\newcommand{\maparrow}[1]{%
\xrightarrow{\mathmakebox[\arrow]{#1}}
}
\newcommand{\map}[3]{%
\mbox{${#2}\!\maparrow{#1}\!{#3}$}
}
\newcommand{\rela}[3]{%
\mbox{${#2}\!\maparrow{#1}\!{#3}$}
}
\newcommand{\revr}[1]{%
{#1}^{r}
}
\newcommand{\compl}[1]{%
\overline{#1}
}
\newcommand{\grap}[1]{%
{#1}^{\#}
}
\newcommand{\rec}[1]{%
\operatorname{Rec}\!\rb{#1}
}
\newcommand{\rat}[1]{%
\operatorname{Rat}\!\rb{#1}
}
\newcommand{\sgwp}[1]{%
\iota_{S}
}
\newcommand{\spr}[2]{%
\operatorname{sg}\!\left<#1~\mid~#2\right>
}
\newcommand{\mpr}[2]{%
\operatorname{mon}\!\left<#1~\mid~#2\right>
}
\newcommand{\rwp}{%
\textbf{rwp}}

\newcommand{\rwps}{%
\textbf{rwp}-semigroup%
}
\newcommand{\gu}[1]{\mathcal{U}\!\left(#1\right)}

\newcommand{\greenR}{\mathcal{R}}
\newcommand{\greenL}{\mathcal{L}}
\newcommand{\greenH}{\mathcal{H}}
\newcommand{\greenJ}{\mathcal{J}}
\newcommand{\greenD}{\mathcal{D}}

\newcommand{\fcs}[1]{%
\operatorname{FCS}\!\rb{#1}
}

\begin{document}


\begin{frontmatter}

\title{Deciding Word Problems of Semigroups using Finite State Automata}

\author[usa]{Max Neunh\"offer}
\ead{max@9hoeffer.de}
\author[usb]{Markus Pfeiffer\corref{cor}}
\ead{markus.pfeiffer@st-andrews.ac.uk}
\author[usa]{Nik Ru\v{s}kuc}
\ead{nik.ruskuc@st-andrews.ac.uk}
\address[usa]{School of Mathematics and Statistics}
\address[usb]{School of Computer Science \\
	 University of St Andrews\\
	 North Haugh \\
	 St Andrews \\
	 KY16 9SS \\
	 Scotland}
\cortext[cor]{Corresponding author, phone: +44 (1334) 46 1633, fax: +44 (1334) 46 3748}

\begin{abstract}
\end{abstract}

\begin{keyword}
  finite state automata \sep rational relations \sep semigroups \sep word problems
\end{keyword}

\end{frontmatter}

\section{Motivation}
\label{sec:motivation}
  When given a semigroup in terms of generators and relations, determining
  whether two given formal products of generators are equal in the semigroup is a
  difficult problem.
  This problem is commonly called the \emph{word problem}, and is
  \emph{undecidable} in general: There are examples of semigroup presentations for
  which there is no algorithm that decides the word problem.

  This is in contrast with semigroups given by concrete generators, such as
  matrices over a ring, or transformations of a finite set.
 
  In this paper we consider a class of semigroups, which we call
  \rwp-semigroups, for which there is a simple algorithm to decide
  the word problem.
  This algorithm runs in constant memory and quadratic time in the length of
  the input.
  Our approach is similar to Sakarovitch's notion of \emph{rational monoids}
  \cite{DBLP:journals/iandc/Sakarovitch87,DBLP:journals/iandc/PelletierS90}, but
  it is not known whether Sakarovitch's rational monoids coincide with
  \rwp-monoids.

  \medskip

  Viewing the word problem as a relation on strings, we review basic definitions
  around semigroups in Section \ref{sec:intro}, define our notion of a
  \emph{generating system} and \emph{word problem} in Section \ref{sec:gensys},
  and the notions of recognisable and rational subsets of monoids in Section
  \ref{sec:rec_and_rat}.

  We then turn our attention to the central notion of interest, \rwp-semigroups
  in Section \ref{sec:rational_word_problem}.
  Since results about word problems are easily generalised to any relation on
  semigroups, we develop the notion of relations with rational word
  problem on finitely generated semigroups as far as possible.
  It is evident that further investigation is desirable in this direction, but
  would exceed the scope of this paper.

  In Section \ref{sec:structure} we show that \rwp-semigroups are closed under
  taking finitely generated subsemigroups, contain an element of infinite order,
  that subgroups of \rwp-semigroups are finite, and more generally that
  $\greenH$-classes are finite, that they have the finiteness properties
  $\greenJ=\greenD$ and residual finiteness, and that Kleene's property holds in
  \rwp-semigroups.
  
  In Section \ref{sec:constr} we show that \rwp-semigroups are closed under
  adding zeros and ones, direct product, if it is finitely generated, extension by
  a finite ideal, semigroup- and monoid-free product, and zero union. 

  In Section \ref{sec:decidability} we discuss decidability results and show
  that it is undecidable whether a finitely presented semigroup is an
  \rwp-semigroup, but once we consider \rwp-semigroups it is decidable whether
  they are finite, a group, a monoid, or free.
  
  We close with a section about further work and open questions.
  
\section{Relations, Strings, and Semigroups}
\label{sec:intro}

  The notions introduced in this section have been studied extensively and
  can be found in literature, for example Berstel's \cite{berstel79}, Eilenberg's
  \cite{eilenbergA}, and Sakarovitch's \cite{sakarovitch2009} monographs.

  \medskip

  We denote the powerset of a set $X$ by $\pows{X}$.
  
  A \emph{relation} $\rela{\rho}{S}{T}$ between sets $X$ and $Y$ is a map
  $\map{\pows{\rho}}{\pows{X}}{\pows{Y}}$ with the property that for any family
  $\rb{X_i}_{i \in I}$ of subsets of $X$ it holds that
  \[
  \rb{\bigcup\limits_{i \in I}X_i}\pows{\rho} = \bigcup\limits_{i\in I}X_i\pows{\rho}.
  \]
  Composition of relations is defined by composition of the underlying
  functions. Note that hence the relation $\rho$ is fully determined by images
  of $\pows{\rho}$ on singleton sets, and hence we denote singleton sets $\{x\}
  \in \pows{X}$ by $x$ when there is no ambiguity.

  For a relation $\rela{\rho}{X}{Y}$ define its graph $\grap{\rho}$ as
  \[
  \grap{\rho} = \cb{ (m,n) \mid n \in m\rho },
  \]
  and the \emph{reverse relation} $\rela{\revr{\rho}}{Y}{X}$ by
  \[
  t\revr{\rho} = \{ x \in X \mid t \in x\rho\}.
  \]

  Note that a function $\map{\varphi}{X}{Y}$ naturally defines a relation
  between $X$ and $Y$, and the relation $\rela{\revr{\varphi}}{Y}{X}$ relates to
  every $y \in Y$ the full preimage of $y$ under the map $\varphi$.

  We define two special relations on every set, 
  the \emph{identity relation} $\rela{\iota_X}{X}{X}$ by $x\iota_X = x$ for every
  $x \in X$, and the \emph{universal relation} $\rela{\mu_X}{X}{X}$ by $x\mu_X = X$
  for every $x \in X$.

  \medskip

  An \emph{alphabet} is a finite set.

  Given an alphabet $A$, a finite sequence of elements from $A$ is called a
  \emph{string} over $A$. We refer to a sequence of length zero as the \emph{empty
  string} and denote it by $\varepsilon_A$, or simply $\varepsilon$ if there is no
  ambiguity. We denote by $\fm{A}$ the set of all strings over $A$, and by
  $\fsg{A}$ the set of all non-empty strings over $A$.

  If $s$ is a string over $A$, we denote by $\abs{s}$ the length of the string, and for
  any $a$ by $\abs{s}_a$ the number of occurrences of the symbol $a$ in $s$.

  We can \emph{concatenate} any two strings $s$ and $t$ over $A$, resulting in a
  new string we denote by $st$. We call $s$ a \emph{prefix} of $t$ if there is a
  string $u$ such that $t = su$, and $s$ a \emph{suffix} of $t$ if there is a
  string $u$ such that $t = us$.

  For any string $s$ over $A$, and any natural number $i$, we denote by $s^i$
  the $i$-fold concatenation of copies of $s$, and set $s^0$ to $\varepsilon_A$.

  \medskip
  
  A \emph{semigroup} is a set $S$ together with a binary associative operation,
  usually denoted $s \cdot t$ or simply $st$ for $s,t \in S$. A \emph{monoid} is a
  semigroup that contains an element $e \in S$ such that for all $s \in S$ it
  holds that $es = se = s$. The element $e$ is usually called the \emph{identity
  element} of the monoid. A \emph{group} is a monoid such that for every element
  $s$ there is an element $t$ such that $st = ts = e$. We call $t$ the
  \emph{inverse} of $s$.

  Any semigroup $S$ can be turned into a monoid by adjoining a new element $e$
  not previously in $S$, and extending the binary operation such that $e$ is the
  new identity element of $S$. We denote the resulting monoid by $\unitadd{S}$.
  We explicitly allow adding a new identity to a monoid.

  An element $z \in S$ is called a \emph{zero} if $zs = sz = z$ for all $s \in S$.
  Just as in the case for adding an identity element, we can adjoin a zero to
  any semigroup $S$ and extend the operation such that $sz = zs = z$ for all $s
  \in S$, and $zz = z$. We denote the result of this construction by $\zeroadd{S}$.

  \medskip
    
  A \emph{semigroup homomorphism} is a map $\map{\varphi}{S}{T}$ from a
  semigroup $S$ to a semigroup $T$ such that $(st)\varphi = (s\varphi)(t\varphi)$.

  \medskip

  Two relevant examples of semigroups and monoids are the sets $\fm{A}$ and
  $\fsg{A}$ with concatenation. The monoid $\fm{A}$ is the \emph{free monoid} on
  $A$ and the semigroup $\fsg{A}$ is the \emph{free semigroup} on $A$.
  We will also briefly be concerned with the \emph{free group} on an alphabet
  $A$, which we denote by $F(A)$.

  The free semigroup $\fsg{A}$ on a set $A$ has the special property that for
  any semigroup $S$ a choice of images for elements of $A$ in $S$ extends uniquely
  to a semigroup homomorphism from $\fsg{A}$ into $S$.
  
  \medskip
  Green's relations form the foundation of any strutural description of
  semigroups, and hence they are described in any standard work on semigroups
  such as \cite{howie1995}.

  For $s$ and $t$ elements of a semigroup $S$, we define the following
  equivalence relations
  \begin{itemize}
    \item $s \greenL t$ if and only if there are $x, y \in \unitadd{S}$ such that
      $xs = t$ and $yt = s$,
    \item $s \greenR t$ if and only if there are $x,y \in \unitadd{S}$ such that 
      $sx = t$ and $ty = s$,
    \item $s \greenH t$ if and only if $s \greenL t$ and $s \greenR t$.
    \item $s \greenD t$ if and only if there is $d \in S$ such that $s \greenL
      d$ and $d \greenR t$.
    \item $s \greenJ t$ if and only if there are $x,y,u,v \in \unitadd{S}$ such that
      $xsy = t$ and $utv = s$.
  \end{itemize}

\section{Generating Systems and Word Problems}
  \label{sec:gensys}

  When defining semigroups in terms of generators and relations, there are
  in general many ways to choose generators, and the representatives of
  elements depend on the choice of generators.

  We consider our choice of generators as ``external'' to the semigroup, because
  we will be talking about formal languages and automata. To make this precise we 
  introduce the notion of a \emph{generating system}.

  \begin{defn}\label{def:gensystem}
    A \emph{generating system} is a map $\map{p}{A}{S}$ where $A$ is a set and
    $S$ is a semigroup.
  \end{defn} 

  If $\map{p}{A}{S}$ is a generating system, then $p$ uniquely extends to a
  semigroup homomorphism $\map{\pi_p}{\fsg{A}}{S}$. In the special situation
  where $\pi_p$ is surjective, we call $p$ a generating system for the semigroup $S$.
  
  \begin{defn}\label{def:gensystem}
    Let $A$ be a set and $S$ be a semigroup. A \emph{generating system for $S$}
    is a map $\map{p}{A}{S}$ that extends to a \emph{surjective} semigroup
    homomorphism $\map{\pi_p}{\fsg{A}}{S}$.
  \end{defn} 

  We call a generating system $\map{p}{A}{S}$ for $S$ \emph{finite} if $A$ is finite,
  and $S$ \emph{finitely generated} if there exists a finite generating system for $S$.
  Note that for any semigroup $S$, the identity map $\map{\iota}{S}{S}$ is a
  generating system.

  Similarly, we can define monoid generating systems for monoids, and group
  generating systems for groups by demanding that the map $p$ extends to a monoid
  homomorphism $\map{\pi_p}{\fm{A}}{M}$, or a group homomorphism
  $\map{\pi_p}{\fg{A}}{M}$

  \medskip

  Given a finite generating system $\map{p}{A}{S}$, the map $\pi_p$ might map
  more than one string to a given element of $S$.
  This raises two questions.

  \begin{enumerate}
  \item\label{t1} Give an algorithm that for the input of two strings $v$
    and $w$ in $\fm{A}$, decides whether $v\pi_p = w\pi_p$.

    If such an algorithm exists, we say \emph{$S$ has decidable word problem}.
  \item\label{t2} Give an algorithm that for the input string $v$ in
    $\fm{A}$ effectively computes a \emph{canonical representative} $c_v$, that
    is $c_v\pi_p=v\pi_p$ and if $w$ is another input with $v\pi_p = w\pi_p$ then
    $c_w = c_v$.
  \end{enumerate}

  Note that a solution to \ref{t2} immediately gives a solution to \ref{t1},
  while a solution to \ref{t1} does not necessarily yield a solution for \ref{t2}.

  A choice of generating system for a semigroup $S$ gives a full description of
  the semigroup as a quotient of the free semigroup $\fsg{A}$ by the kernel of
  $\pi_p$.

  \begin{defn}
    Let $S$ be a semigroup and let $\map{p}{A}{S}$ be a finite generating system
    for $S$.
    The \emph{semigroup word problem} of $S$ with respect to the generating
    system $p$ is the relation $\pi_p\revr{\pi_p}$, which we denote by
    $\iota\!\rb{p}$.
  \end{defn}
  Note that
  \[
   \grap{\iota\!\rb{p}}
    = \cb{ (v,w) \in \fsg{A}\times\fsg{A} ~\mid~ v\pi_p = w\pi_p }.
  \]

  \medskip

  Since $\iota\!\rb{p} = \pi_p\iota_S\revr{\pi_p}$ it is natural to
  generalise the above definition by replacing the identity relation by just any
  relation, and introduce the notion of word problem of a relation.

  \begin{defn}
    Let $S$ be a semigroup, $\map{p}{A}{S}$ be a finite generating system for $S$,
    and let $\rela{\rho}{S}{S}$ be a relation on $S$.
    Then the \emph{word problem of $\rho$ with respect to the generating system $p$}
    is the relation $\pi_p\rho\revr{\pi}_p$, denoted by $\rho\!\rb{p}$.
  \end{defn}

  Note that, slighty abusing names, we call the word problem of
  $\iota_S\!\rb{p}$ the word problem of the semigroup $S$.

  The two questions above can thus be rephrased as
  \begin{itemize}
  \item[i)] Give an algorithm that for the input of two strings $v$
    and $w$ in $\fm{A}$, decides whether $v \in w\rho\!\rb{p}$.
  \item[ii)] If $\rho$ is an equivalence relation, give an algorithm
    that for the input of string $v$ in $\fm{A}$ effectively computes a
    canonical representative for the equivalence class of $v$.
  \end{itemize}

\section{Recognisable and Rational}
  \label{sec:rec_and_rat}

  Taken at full generality the definitions of Section \ref{sec:gensys} are not of
  great use: canonical representatives are not computable and word problems are
  undecidable.

  We want to look at some well-defined classes of semigroups in which the word
  problem is not only decidable, but efficiently so.
  To this end, we collect some definitions and results which are
  well-examined in literature, for example in Eilenberg's \cite{eilenbergA}
  and Sakarovitch's \cite{sakarovitch2009} monographs.
  
  We give definitions of \emph{recognisable} and \emph{rational} subsets of
  semigroups and monoids. In the case of recognisable and rational subsets of the
  free semigroup or monoid there is a strong connection to computation using
  finite state automata.



  \medskip
  In what follows let $S$ be a semigroup.
  \medskip

  \begin{defn}
    A subset $X \subset S$ is called \emph{recognisable} if there is a semigroup
    homomorphism $\map{\varphi}{S}{T}$ such that $T$ is finite, and
    $X\varphi\varphi^{-1} = X$. We denote the family of recognisable subsets of $S$
    by $\rec{S}$.
  \end{defn}

  \medskip


  The following auxiliary definitions are necessary to give a definition of the
  family of rational subsets of $S$.
  Let $X$ and $Y$ be subsets of $S$, then
  \[
    XY = \cb{ xy \mid x \in X, y \in Y }
  \]
  is a subset of $S$. For any $n \in \N$, define subsets
  \[
    X^{n+1} = XX^n,
  \]
  and the \emph{Kleene plus $X^+$} of $X$ to be 
  \[
    X^+ = \bigcup\limits_{n \in \N_{+}}X^n.
  \]

  If $S$ is a monoid, we can also define $X^0 = \{ e \}$ and $X^* = \bigcup\limits_{n \in \N}X^n$.



  \begin{defn}
    \label{defn:rationalsubsets}
    The family $\rat{S} \subset \pows{S}$ of \emph{rational subsets} of $S$ is
    inductively defined:
    \begin{itemize}
    \item any finite subset of $S$ is in $\rat{S}$, and
    \item if $X \in \rat{S}$ and $Y \in \rat{S}$, then $X^+ \in \rat{S}$, $X
      \cup Y \in \rat{S}$ and $XY \in \rat{S}$.
    \end{itemize} 
  \end{defn}

  Definition \ref{defn:rationalsubsets} allows specifying rational subsets of
  semigroups by giving rational expressions built using the operations used.

  McKnight \cite{mcknight1964} shows that, if $S$ has a finite generating
  system, then $\rec{S} \subset \rat{S}$. Kleene's theorem \cite{kleene56} states
  that for {$S=\fsg{A}$} the families of recognisable and rational subsets
  coincide.

  The converse of Kleene's theorem does not hold in general, and we call
  semigroups in which it does \emph{Kleene semigroups}.

  There is no known characterisation of Kleene semigroups.

  
  In some parts of the literature, recognisable subsets of free monoids and
  semigroups are better known as \emph{regular languages}, which are often
  introduced as languages that can be recognised by finite state automata.


  If we intersect a rational set with a recognisable one, the result is
  still rational, for a proof see \cite[Ch 2, Prop 2.6]{sakarovitch2009}.
  \begin{prop}\label{prop:rec_rat_intersection}
    If $X \subset S$ is rational and $Y \subset S$ is recognisable,
    then $X \cap Y$ is rational.
  \end{prop}

  The following \emph{pumping lemma} provides a sufficient condition for
  subsets of $\fsg{A}$ to be recognisable, and hence is a useful tool to prove
  non-recognisablity.
  For a proof see \cite[Ch I, Lemma 1.3]{sakarovitch2009}
  \cite[Ch IV, Propsition 1.6]{sakarovitch2009}
  \begin{prop}\label{prop:pumping_recognisable}
    Let $X \subset A^+$ be recognisable. Then there exists $n_X \in \mathbb{N}$ such
    that for every $v \in X$ with $\abs{v} > n_X$ there is a factorisation $v = xyz$
    with $\abs{xy} \leq n_X$ such that $xy^iz \in X$ for all $i \in \mathbb{N}$.
  \end{prop}

  \section{Rational (Equivalence) Relations}
  
  We call a relation $\rela{\rho}{S}{T}$ \emph{recognisable} if its graph is a
  recognisable subset of $S \times T$, and \emph{rational} if its graph is a
  rational subset of $S \times T$.

  \medskip

  We will use the following propositions in our work, proofs can for example be
  found in \cite{berstel79, eilenbergA, sakarovitch2009}.

  Rational relations are closed under composition, given that the middle semigroup
  is free, see \cite[Ch IV, Thm 1.4]{sakarovitch2009}.
  \begin{prop}
    \label{prop:rat_closed_composition}
    If $\rela{\rho}{S}{\fm{B}}$ and $\rela{\sigma}{\fm{B}}{T}$ are rational
    relations, then $\rela{\rho\sigma}{S}{T}$ is rational.
  \end{prop}

  Homomorphisms between free semigroups and monoids are rational, cf
  \cite[Ch IV, Examples 1.1]{sakarovitch2009}.

  \begin{prop}
    \label{prop:free_homomorphism}
    If $A$ and $B$ are finite sets, then any homomorphism $\map{\varphi}{\fm{A}}{\fm{B}}$ is
    rational.
    \begin{proof}
      Let $\map{\varphi}{\fm{A}}{\fm{B}}$ be a homomorphism. Then a rational
      expression for its graph is
      \[
        \rb{\bigcup\limits_{a \in A}\rb{a,a\varphi}}^*.
      \]
    \end{proof}
  \end{prop}

  The following \emph{pumping lemma} provides a sufficient condition for
  relations $\rela{\rho}{\fsg{A}}{\fsg{A}}$ to be recognisable, and hence is a
  useful tool to prove non-rationality. For a proof see
  \cite[Ch IV, Propsition 1.6]{sakarovitch2009}, and also compare with
  Proposition~\ref{prop:pumping_recognisable}.
  
  \begin{prop}\label{prop:pumping_rational}
    Let $\rela{\rho}{\fsg{A}}{\fsg{A}}$ be a rational relation. Then there exists
    $n_\rho \in \mathbb{N}$ such that whenever $w \in v\rho$ with $\abs{v} + \abs{w} > n_\rho$
    there are factorisations $v=x_vy_vz_v$ and $w = x_wy_wz_w$ with
    $\abs{x_vy_v} + \abs{x_wy_w} \leq n_\rho$ such that $x_wy_w^iz_w \in (x_vy_v^iz_v)\rho$ for
    all $i \in \mathbb{N}$.
  \end{prop}

  For rational equivalence relations some interesting results are already known,
  and most of them are due to Johnson \cite{johnsonphd}, but some earlier 
  results are due to Elgot and Mezei \cite{Elgot:1965:RDG:1664570.1664574}.
  These results will feed into results about rational word problems.

  \begin{prop}\label{prop:recog_finite}
    An equivalence relation $\rela{\rho}{\fm{A}}{\fm{A}}$ is recognisable if and only if
    it has finitely many equivalence classes.
  \end{prop}

  Johnson also proves in \cite{johnsonphd} that for every rational equivalence
  relation there is a recognisable set of strings such that its intersection with
  every equivalence class is finite. Note that there is no bound on the size of
  the intersection.

  \begin{prop}\label{prop:rational_finite_rep}
    If $\rela{\rho}{\fm{A}}{\fm{A}}$ is a rational equivalence relation, then there
    exists a recognisable subset $D \subset \fm{A}$ such that $v\rho \cap D$ is
    non-empty and finite for all $v \in \fm{A}$.
  \end{prop}

  It is a long-standing open question whether there exists a recognisable set
  of \emph{unique} representatives for any given rational equivalence relation.
  \begin{que}
    \label{que:ratcross}
    Do rational equivalence relations have rational cross-sections?
  \end{que}

%

\section{Rational Word Problem}

  \label{sec:rational_word_problem}
  Taking the preceding sections together we introduce the central concepts of
  this paper: relations on semigroups that have recognisable or rational word
  problem.

  Since the definition of \emph{rational word problem} as it stands depends on
  the choice of generating system, we first show that having rational word
  problem is independent of the choice of generating system.
  This justifies saying that a relation has \emph{rational word problem} without
  mentioning a generating system.


  \begin{thm}
    \label{thm:changens}
    Let $S$ be a semigroup, and let $\map{\rho}{S}{S}$ be a relation on $S$.

    If there is a finite generating system $\map{p}{A}{S}$ \emph{for $S$} such that
    the word problem $\rho\!\rb{p}$ is rational, then for any finite
    generating system $\map{q}{B}{S}$ the word problem $\rho\!\rb{q}$ is also
    rational.

    \begin{proof}
      Let $S, \rho, p,$ and $q$ as stipulated in the hypothesis.

      Choose
      
      for each $b \in B$ an element $v_b \in b\pi_q\revr{\pi_p}$, and
      let $\map{\varphi}{\fm{B}}{\fm{A}}$ be the unique homomorphism defined by
      by $b\varphi = v_b$. Note that for all $v \in \fm{B}$ it holds that
      $v\pi_q = v\varphi\pi_p$, and that by Proposition \ref{prop:free_homomorphism}
      $\varphi$ defines a rational relation.

      If we show that
      \[
      \rho\!\rb{q} = \varphi\rho\!\rb{p}\revr{\varphi},
      \]
      we can conclude that $\rho\!\rb{q}$ is rational as a composition of
      rational relations. 

      Let $v,w \in \fm{B}$, then
      \begin{align*}
          w \in v\rho\!\rb{q} & \Leftrightarrow w\pi_q \in v\pi_q\rho\\
          & \Leftrightarrow w\rb{\varphi\pi_p} \in v\rb{\varphi\pi_p}\rho\\
          & \Leftrightarrow w \in v\rb{\varphi\pi_p}\rho\revr{\rb{\varphi\pi_p}}\\
          & \Leftrightarrow w \in v\rb{\varphi\rho\!\rb{p}\revr{\varphi}}.
      \end{align*}
    \end{proof}
  \end{thm}

  An immediate consequence of Theorem \ref{thm:changens} is that rationality of
  the word problem of a relation does not depend on the choice of a generating system.

  
  Theorem \ref{thm:changens} justifies the following definition.

  \begin{defn}
    Let $\rela{\rho}{S}{S}$ be a relation on a semigroup $S$.
    If there exists a finite generating system $\map{p}{A}{S}$ such that
    $\rho\!\rb{p}$ is rational, we call $\rho$ an \emph{\textbf{rwp}-relation}
    or say that $\rho$ has \emph{rational word problem}.
  \end{defn}

  In this paper we will consider semigroups where $\iota_S\!\rb{p}$ is rational,
  and call such semigroups \textbf{rwp}-semigroups.
  
  \begin{defn}
    A semigroup $S$ is an \rwp-semigroup if $\rela{\iota_S}{S}{S}$ has rational word problem.
  \end{defn}

  \subsection{Examples}
  \label{sec:examples}

  In this section we exhibit some examples and non-examples of \rwp-relations,
  and \rwp-semigroups.

  \medskip
  
  For any finitely generated semigroup $S$ the universal relation
  $\rela{\mu_S}{S}{S}$ is an \rwp-relation, and for any finitely
  generated group $G$ the all Green's relations are \rwp-relations.

  \begin{exam}
    The free semigroup $\fsg{A}$ has rational word problem. Let
    $\map{p}{A}{\fsg{A}}$ map $a \in A$ to the string $a$ of length $1$.

    The relation $\iota_{\fsg{A}}\!(p)$ is rational, because it can be given as
    \[
      \left(\bigcup\limits_{a \in A} (a,a)\right)^+.
    \]
  \end{exam}

  \begin{exam}
    The semigroup $E = \spr{a,b}{a^2=a, ba=a}$ is an \rwp-semigroup. Consider the generating system
    $\map{p}{\cb{a,b}}{S}$. We claim that
    \begin{equation}
      \rb{b,b}^+ \cup \cb{\rb{a,\varepsilon},\rb{b,\varepsilon},\rb{\varepsilon,a},\rb{\varepsilon,b}}^*\rb{a,a}\rb{b,b}^* \label{wpreleq}
    \end{equation}
    is the graph of $\iota_E\!\rb{p}$.
    
    Observe that $(v,w)$ is in the relation given in equation (\ref{wpreleq}) if an only if either
    $v=w=b^k$ for some $k \in \N \backslash \{0\}$, or $v = v'ab^k$ and $w = w'ab^k$ for some $v', w' \in \fm{\{a,b\}}$
    and some $k \in \N$.

    For $v=v'ab^k$ we claim that $v\iota_E\!\rb{p}~=~\{ wab^k ~|~ w \in \fm{\{a,b\}} \}$.

    Let $v=v'ab^k$ for some $k\in \N$, then using the equations $a^2 = a$ and
    $ba = a$ we can rewrite $v$ to $ab^k$, and then rewrite to any $w \in
    \fm{\{a,b\}}$.
  \end{exam}

  \begin{exam}
    The semigroup $F = \spr{a,b}{ab^na = aba}$ is an \rwp-semigroup since the graph
    of the word problem is
    \[
      \rb{\rb{\varepsilon,\varepsilon} \cup \rb{b,b}^+}\rb{a,a}^+\rb{b,b}\rb{\rb{b,\varepsilon}, \rb{\varepsilon,b}}^+\rb{a,a}^+.
    \]
    We omit the proof because it is very similar to the one for $E$, but note that
    $F$ does not have a finite presentation.
  \end{exam}
  
  \begin{exam}
    We show that the free commutative semigroup
    $\fcs{a,b} = \spr{a,b}{ab=ba}$
    is \emph{not} an \rwp-semigroup.

    First, observe that $w \in v\iota_{\fcs{a,b}}(\{a,b\})$ if and only if $\abs{v}_a =
    \abs{w}_a$ and $\abs{v}_b = \abs{w}_b$.

    For a contradiction assume that $\iota_{\fcs{a,b}}\!\rb{A}$ is rational, hence by
    Proposition \ref{prop:pumping_rational} there exists $n_{\iota} \in \N$
    such that whenever $w \in v\iota_S\!\rb{A}$ with
    $|v| + |w| > n_{\iota}$, there are factorisations
    \[
      v=v_1v_2v_3 \mbox{ and } w = w_1w_2w_3
    \]
    with $|v_1v_2| + |w_1w_2| < n_{\iota}$, and such that
    $w_1w_2^iw_3 \in \rb{v_1v_2^iv_3}\iota_{\fcs{a,b}}\!\rb{A}$ for all $i \in \N$.
    
    For $n > n_\iota$ consider
    \[
      v = a^nb^n \mbox{ and } w = b^na^n.
    \]
    Proposition \ref{prop:pumping_rational} ensures that there are
    $x_1,x_2,x_3,y_1,y_2,y_3 \in \N$ with $n = x_1 + x_2 + x_3 = y_1 + y_2 + y_3$
    and $x_1 + x_2 + y_1 + y_2 < n_\iota$ such that
    \[
      v = a^{x_1}a^{x_2}a^{x_3}b^n \mbox{ and }  w = b^{y_1}b^{y_2}b^{y_3}a^n,
    \]
    and such that $b^{y_1}b^{iy_2}b^{y_3}a^n \in
    \rb{a^{x_1}a^{ix_2}a^{x_3}b^n}\iota_S\!\rb{A}$ for all $i \in \N$, a
    contradiction, because $x_1 + 2x_2 + x_3 > x_1 + x_2 + x_3$.
  \end{exam}

  We also remark that Sakarovitch's rational monoids \cite{DBLP:journals/iandc/Sakarovitch87} are
  \rwp-monoids. We first give a definition of a rational monoid.
  \begin{defn}
    \label{def:rational}
    Let $S$ be a monoid. We call $S$ rational if there is a finite generating
    system $\map{p}{A}{S}$, and an injective map $\map{u}{S}{\fm{A}}$ such that 
    $\pi_p\circ u$ is rational.
  \end{defn}

  The image of $u$ is a rational cross section, that is a set of unique
  representatives of elements of $S$.
  An answer to Question \ref{que:ratcross} would establish whether \rwp-monoids
  are rational.
  
  \begin{thm}
    Let $S$ be a rational monoid. Then $S$ is an \rwp-monoid.
    \begin{proof}
      Let $\map{p}{A}{S}$ be a finite generating system and $\map{u}{S}{\fm{A}}$
      the representative map that exists by Definition \ref{def:rational}. Then
      the relation $u\revr{u} = \iota_S$, and hence
      \[
        (\pi_p \circ u)\circ\revr{(\pi_p\circ u)}
        = \pi_p \circ (u \circ \revr{u}) \circ \revr{\pi_p}
        = \pi_p \circ \iota_S\circ\revr{\pi_p} 
      \]
      is rational and the word problem of $S$ with respect to the generating
      system $p$.
    \end{proof}
  \end{thm}

\section{Structure}
  \label{sec:structure}

  In this section we explore structural properties of \rwp-semigroups.

  \medskip
  
  For completeness, we briefly consider semigroups with \emph{recognisable} word
  problem. Anisimov showed in \cite{grouplanguages1971} that a group has
  recognisable word problem if and only if it is finite. The following result is
  the semigroup analogue.

  In group theory the notion of coword problem, the complement of the word
  problem, has sparked considerable interest, and it seems natural to consider
  coword problems in our setting too. We only include results where they follow
  directly and defer further investigation to later papers.

  \begin{thm}
    Let $S$ be a semigroup and $\map{p}{A}{S}$ a finite generating system.

    The following are equivalent.
    \begin{enumerate}
    \item $S$ is finite
    \item $\iota_S\!\rb{p}$ is recognisable
    \item the complement $\compl{\iota_S\!\rb{p}}$ of $\iota_S\!\rb{p}$ is
      recognisable.
    \end{enumerate}
    \begin{proof}
      If $S$ is a finite semigroup, then $\unitadd{S} \times \unitadd{S}$
      is a finite semigroup and
      \[
      \varphi : \fsg{A} \times \fsg{A} \rightarrow \unitadd{S} \times \unitadd{S}, (v,w) \mapsto (v\pi_p,w\pi_p)
      \]
      is a semigroup homomorphism.

      Consider the graph $G$ of $\iota_S\!\rb{p}$. It holds that
      ${G\varphi\varphi^{-1} \subset G}$. To show equality, let $(v,w) \in G$,
      then $(v,w)\varphi = (v\pi_p,w\pi_p) = (s,s)$ for some $s \in S$,
      hence ${(v,w) \in G\varphi\varphi^{-1}}$, and therefore $\iota_S\!\rb{p}$ is
      recognisable.

      If $\iota_S\!\rb{p}$ is recognisable, it has finitely many equivalence
      classes by Proposition \ref{prop:recog_finite}, hence $S$ is finite.
      
      Furthermore, the recognisable subsets of a semigroup form a Boolean
      algebra, therefore the complement of $\iota_S\!\rb{p}$ is recognisable
      if and only if $\iota_S\!\rb{p}$ is.
    \end{proof}
  \end{thm}

  As a direct consequence of Theorem \ref{thm:changens}, finitely generated
  subsemigroups of \rwp-semigroups are \rwp-semigroups.

  \begin{thm}\label{thm:finitely_generated_subsemigroups}
    \label{thm:subsemis}
    If $S$ is an \rwp-semigroup, and $T \subset S$ is a finitely generated subsemigroup, then
    $T$ is an \rwp-semigroup.
    \begin{proof}
      This follows directly from Theorem \ref{thm:changens} using $\rela{\iota}{S}{S}$ and
      any finite generating set $\map{p}{A}{S}$ and a finite generating set $\map{q}{B}{S}$ for $T$.
    \end{proof}
  \end{thm}

  And as a corollary an \rwp-semigroup cannot contain a finitely generated
  non-\rwp-semigroup, such as a free commutative semigroup.

  \begin{cor}
    If $S$ is a finitely generated semigroup that contains a free commutative
    semigroup of rank at least two, then $S$ is not an \rwp-semigroup.
  \end{cor}

  Any infinite \rwp-semigroup contains an infinite monogenic subsemigroup.
  
  \begin{thm}\label{thm:infinite_order_element}
    Let $S$ be an infinite \rwp-semigroup. There exists an $s \in S$ such that the
    subsemigroup generated by $s$ is infinite.
    \begin{proof}
      Let $S$ be an infinite \rwp-semigroup, and $\map{p}{A}{S}$ a finite generating
      system for $S$. By Proposition \ref{prop:rational_finite_rep} there exists a
      recognisable subset $D$ of $\fsg{A}$ such that $0 < \abs{s\pi_p^{-1} \cap D} < \infty$.
      for any $s \in S$.

      We apply Proposition \ref{prop:pumping_recognisable} for recognisable
      subsets of $\fsg{A}$: Let $v \in D$ with $\abs{v} > n_D$, thus
      there are $x, y, z \in \fsg{A}$ such that $v = xyz$, with $\abs{y} < N_D$
      and such that $xy^iz \in D$ for all $i \in \mathbb{N}$.

      This means that the set
      $R = \{ (xy^iz)\pi_p ~|~ i \in \mathbb{N} \}$ is an infinite subset of $S$.

      The element $s = y\pi_p$ generates an infinite subsemigroup of $S$; suppose
      that to the contrary there exist $i,j \in \mathbb{N}$ such that $s^i = s^j$,
      then the set $R$ would be finite. 
    \end{proof}
  \end{thm}

  \subsection{Green's Relations and Subgroups}
  \label{sec:green}

  We cover two aspects of Green's relations on \rwp-semigroups in this section:
  finiteness of $\greenH$ classes, and the fact that $\greenJ = \greenD$.

  Let $H$ be an $\greenH$-class of $S$, and {$T_H = \{ t \in \unitadd{S} ~|~ Ht
    \subset H\}$}. For every $t \in T_H$ we define a transformation
  $\map{\gamma_t}{H}{H}$ by $h\gamma{t} = ht$. One can show that every $\gamma_t$
  is invertible, and hence the set of all $\gamma_t$ forms a permutation group
  under composition of maps. This group is known as the \emph{Sch\"utzenberger
    group of $H$}. If $H$ contains an idempotent then the Schützenberger group of
  $H$ is isomorphic to $H$. For details and proof we refer the reader to
  \cite{clifprestv1}.

  We show that Schützenberger groups of $H$ classes of \rwp-semigroups are finite.

  \begin{thm}
    Let $S$ be an \rwp-semigroup. Then every $\greenH$ class of $S$ is finite.
    \begin{proof}
      Let $S$ be an \rwp-semigroup and let $\map{p}{A}{S}$ be a finite generating
      set for $S$. Let $n_{\iota_S\!(p)}$ be the constant guaranteed to exist by Proposition
      \ref{prop:pumping_recognisable}.
      Assume for a contradiction that $S$ has an infinite $\greenH$-class $H$. Choose
      $h \in H$, and $v \in \fsg{A}$ with $v\pi_p = h$. Since $H$ is assumed to be infinite
      the Schützenberger group $\mathcal{S}\!(H)$ of $H$ is an infinite transformation group.

      There exists $g \in \mathcal{S}\!(H)$ represented by $w \in \fsg{A}$ such that the
      shortest string $w' \in \fsg{A}$ with $(vww')\pi_p = h$ satisfies
      $\abs{w'} > (|v| + 1)n_{\iota_S\!(p)} + |v|$.

      It holds that $vww' \in v\iota_S\!(p)$, and that $vwxu^iy \in v\iota_S\!(p)$, so
      $vwxy \in v\iota_S\!(p)$ contradicting the choice of $w'$ to be of minimal length.
    \end{proof}
  \end{thm}

  \begin{cor}
    \label{cor:groupsfinite}
    Every subgroup of an \rwp-semigroup is finite.
  \end{cor}

  For any semigroup it holds that $\greenJ \subset \greenD$, whereas equality of
  $\greenJ$ and $\greenD$ is a commonly examined finiteness condition.
  We show that \rwp-semigroups have this property.

  \begin{lem}
    If $S$ is an \rwp-semigroup then for any $a,b \in \unitadd{S}$ it holds that
    if $a\unitadd{S} \subset ba\unitadd{S}S$ implies $a\unitadd{S} = ba\unitadd{S}$.
    \begin{proof}
      Let $T = \unitadd{S}$, which is an \rwp-semigroup by Theorem \ref{thm:add_elements}.
      
      We first show that for any $a, b \in T$ with $aT \subset baT$ implies
      $aT = baT$.

      If $aT \subset baT$ then $TaT \subset TbaT \subset TaT$ and so
      $a\greenJ = ba\greenJ$.

      Choose $v$ and $w$ from $\fm{A}$ such that $v\pi = a$ and $w\pi=b$. There
      exist $x,y$, and $z$ in $\fm{A}$ such that $(wy)\pi = a$ and $(xvz)\pi =
      ba$. We can hence replace any occurrence of $v$ by $wy$ and any occurrence
      of $w$ by $xvz$, therefore
      \[
        ba = (xvz)\pi = (xwyz)\pi = (xxvzyz)\pi = \ldots
      \]
      and it follows that $(x^nw(yz)^n)\pi = ba$, and so $x^nw(yz)^n \in
      w\iota_S(p)$ for every $n \in \N_{>0}$

      This implies $ba \in aY$, and therefore $aT=baT$. An analogous argument
      can be used to show that $Ta = Tba$.
    \end{proof}
  \end{lem}
  
  \begin{thm}
    If $S$ is an \rwp-semigroup, then $\greenJ = \greenD$.
    \begin{proof}
      Let $T = \unitadd{S}$.

      Note that in every semigroup it holds that $\greenD \subset \greenJ$.

      Let now $a\greenJ = b\greenJ$, that is, there exist $x$ and $z$ in $T$
      with $a = xbz$, and the goal is to show that $a\greenD=b\greenD$, that is, the
      existence of $c\in T$ such that $a\greenR=c\greenR$ and $c\greenL=b\greenL$.

      Note that due to $a\greenJ=b\greenJ$ it holds that $Ta \subset Tb$ implies
      $Ta = Tb$ and $aT \subset bT$ implies $aT = bT$.
      
      We choose $c = xb$, and remark that $xb\greenJ = b\greenJ$,
      and $Txb \subset Tb$, hence $Txb=Tb$, and so $xb\greenL=b\greenL$.

      Furthermore,$aT = xbzT \subset xbT$, and therefore $aT = xbT$, that is,
      $a\greenR=xb\greenR$.

      We have hence shown that $\greenJ=\greenD$.
    \end{proof}
  \end{thm}

  \subsection{Kleene's Theorem}
  \label{sec:kleenes_theorem}

  We show that in any \rwp-semigroup $S$ it holds that $\rat{S} = \rec{S}$, that
  is \rwp-semigroups are Kleene semigroups, as a corollary we show that
  \rwp-semigroups are residually finite.

  To this end we first show that the preimage of a rational subset of $S$ is
  a rational subset of $\fsg{A}$.
  
  \begin{lem}\label{lem:preim_rational_rational}
    Let $S$ be an \rwp-semigroup and $\map{p}{A}{S}$ be a generating system
    for $S$. If $X \subset S$ is rational, then the preimage
    $X\pi_p^{-1} \subset \fsg{A}$ is a rational.
    \begin{proof}
      We proceed by induction.
      \begin{itemize}
        \item Let ${s} \subset S$ be a singleton subset of $S$, and let 
          $v \in S$ with $v\pi_p = s$. Since $\iota_S\!(p)$ is rational,
          $v\iota_S\!(p) \subset \fsg{A}$ is rational. Therefore preimages
          of finite subsets of $S$ are rational subsets of $\fsg{A}$.

        \item Let $Z = X \cup Y$, where $X$ and $Y$ are rational subsets of
          $S$, and by induction $X\pi_p^{-1}$ and $Y\pi_p^{-1}$ are rational
          subsets of $\fsg{A}$. Then
          \[
          Z\pi_p^{-1} = \left(X\cup Y\right)\pi_p^{-1} = X\pi_p^{-1} \cup Y\pi_p^{-1}
          \]
          is rational.
      
        \item Let $Z = XY$, where $X$ and $Y$ are rational subsets of $S$
          and by induction $X\pi_p^{-1}$ and $Y\pi_p^{-1}$ are rational
          subsets of $\fsg{A}$. Then
          \[
          Z\pi_p^{-1} = (XY)\pi_p^{-1} = (X\pi_p^{-1})(Y\pi_p^{-1})
          \]
          is rational, because
          \begin{align*}
            z \in (XY)\pi_p^{-1} & \Leftrightarrow z\pi_p = xy \mbox{ for $x \in X$ and $y \in Y$}\\
             & \Leftrightarrow z\pi_p = (v\pi_p)(w\pi_p) \mbox{ for $v \in X\pi_p^{-1}$ and $w \in Y\pi_p^{-1}$}\\
            & \Leftrightarrow vw \in z\iota_S\!(p) \mbox{ for $vw \in (X\pi_p^{-1})(Y\pi_p^{-1})$}
          \end{align*}
        \item Let $Z = X^+$, where $X$ is a rational subset of $S$ such that
          $X\pi_p^{-1}$ is a rational subset of $\fsg{A}$. Then
          \[
            Z\pi_p^{-1} = (X^+)\pi_p^{-1} = (X\pi_p^{-1})^+
          \]
          is a rational subset of $\fsg{A}$, because
          \begin{align*}
            z \in (X^+)\pi_p^{-1}&\Leftrightarrow z\pi_p = x_1x_2\cdots x_n\mbox{ for  $x_i \in X$}\\
            &\Leftrightarrow v \in z\iota_S\!(p) \mbox{ for $v \in \left(X\pi_p^{-1}\right)^+$}
          \end{align*}
       \end{itemize}
    \end{proof}
  \end{lem}

  \begin{thm}\label{thm:reg_rat}
    Let $S$ be a semigroup and $\map{p}{A}{S}$ a finite generating system.
    $X \subset S$ is recognisable if and only if $X\pi_p^{-1}$ is rational.

    \begin{proof}
      If $X$ is a recognisable subset of $S$, then $X\pi_p^{-1}$ is a recognisable subset of $\fsg{A}$,
      and by Kleene's theorem $X\pi_p^{-1}$ is a rational subset of $\fsg{A}$.

      Conversely let $X \subset S$ such that $X\pi_p^{-1}$ is rational. By Kleene's theorem
      there exists a morphism $\map{\varphi_X}{\fsg{A}}{T_X}$ such that $T_X$ is finite and
      $X\pi_p^{-1} = X\pi_p^{-1}\varphi_X\varphi_X^{-1}$.

      \begin{align*}
        w \in v\ker\pi_p &\Rightarrow v\pi_p = w\pi_p\\
        & \Rightarrow (\forall x,y \in \fm{A}) (xvy)\pi_p = (xwy)\pi_p\\
        & \Rightarrow (\forall x,y \in \fm{A}) (xvy)\pi_p\pi_p^{-1} = (xwy)\pi_p\pi_p^{-1}\\
        & \Rightarrow (\forall x,y \in \fm{A}) xvy \in X\pi_p^{-1} \Leftrightarrow xwy \in X\pi_p^{-1}\\
        & \Rightarrow w \in v\ker\varphi_X
      \end{align*}
    \end{proof}
  \end{thm}

  \begin{cor}
    Let $S$ be a semigroup with finite generating system $\map{p}{A}{S}$.

    If every for every $X \subset S$ that is rational, the preimage
    $X\pi_p^{-1}$ is rational, then $S$ is a Kleene semigroup.
    \begin{proof}
      This follows directly from \ref{thm:reg_rat}
    \end{proof}
  \end{cor}

  In conclusion we show that \rwp-semigroups are Kleene.
  \begin{thm}\label{thm:rat_implies_kleene}
    Every \rwp-semigroup is a Kleene semigroup.
    \begin{proof}
      Let $S$ be an \rwp-semigroup. Since $S$ is finitely generated,
      McKnight's result \cite{mcknight1964}  ensures that $\rec{S} \subset
      \rat{S}$.
      The claim now follows from Lemma \ref{lem:preim_rational_rational} and
      Theorem \ref{thm:reg_rat}.
    \end{proof}
  \end{thm}

  A semigroup $S$ is \emph{residually finite} if for any two distinct elements
  $s$ and $t$ in $S$ there is a semigroup homomorphism $\map{\varphi}{S}{T}$
  such that $T$ is finite and $s\varphi \neq t\varphi$.
  
  A consequence of Theorem \ref{thm:rat_implies_kleene} is that
  \rwp-semigroups are residually finite.

  \begin{cor}
    Every \rwp-semigroup is residually finite.
    \begin{proof}
      Let $s$ and $t$ be distinct elements of an \rwp-semigroup $S$. Then
      the set $\{s\}$ is a rational subset of $S$, and by Theorem \ref{thm:rat_implies_kleene}
      a recognisable subset of $S$, hence there is a homomorphism
      $\map{\varphi_s}{S}{T}$ with $T$ finite such that
      $\{s\}\varphi_s\varphi_s^{-1} = \{s\}$, hence $s\varphi_s \neq t\varphi_s$.
    \end{proof}
  \end{cor}

\section{Constructions}
  \label{sec:constr}

  This section is dedicated to showing closure of \rwp-semigroups under some
  standard algebraic constructions.
  We consider adding zeros and ones, or a finite ideal, forming a direct
  product, free product, and zero union, and find that \rwp semigroups are
  closed under all these constructions. In the case of direct products we have
  the condition that the product is finitely generated, and in the case of
  monoid-free products only one of the factors can have a non-trivial group of
  units.

  \subsection{Adding zeros and ones}
  \begin{thm}\label{thm:add_elements}
    Let $S$ be a semigroup. Then the following are equivalent.
    \begin{enumerate}
    \item $S$ is an \rwp-semigroup
    \item $\zeroadd{S}$ is an \rwp-semigroup
    \item $\unitadd{S}$ is an \rwp-semigroup
    \end{enumerate}
    \begin{proof}

      If $\zeroadd{S}$ or $\unitadd{S}$ is an \rwp-semigroup, then $S$ is a
      finitely generated subsemigroup of $\zeroadd{S}$ or $\unitadd{S}$
      respectively, and hence an \rwp-semigroup by
      \ref{thm:finitely_generated_subsemigroups} $S$ is an \rwp-semigroup.

      Let $S$ be an \rwp-semigroup and let $\map{p}{A}{S}$ be a generating system for $S$.
      Let $B = A \cup \{b\}$, where $b$ is a new symbol not in $A$.

      Consider $\map{\zeroadd{p}}{B}{\zeroadd{S}}$ with $\zeroadd{p}(x) = p(x)$ for all $x \in A$
      and $\zeroadd{p}(b) = z$. This is a finite generating system for $\zeroadd{S}$, and
      \[
        \iota_{\zeroadd{S}}(\zeroadd{p}) = \iota_S(p) \cup \mu_{\fm{B}b\fm{B}}
      \]
      which is rational as a union of rational relations.
      
      Consider $\map{\unitadd{p}}{B}{\unitadd{S}}$ with $\unitadd{p}(x) = p(x)$
      for all $x \in A$ and $\unitadd{p}(b) = e$ is a finite generating system
      for $\unitadd{S}$.
      There is a unique homomorpism $\map{\varphi}{\fsg{B}}{\fm{A}}$
      defined by $x\varphi = x$ for $x \in A$ and $b\varphi = \varepsilon$.

      Since $\map{\iota_S\!\rb{p}}{\fsg{A}}{\fsg{A}}$ is rational,
      $\map{\iota_S\!\rb{p}}{\fm{A}}{\fm{A}}$ is rational, and so
      $\iota_S\!\rb{p^*} \cup \{(\varepsilon,\varepsilon)\}$
      is rational.
      
      Then
      \[
        \iota_{\unitadd{S}}(\unitadd{p}) = \varphi \circ \iota_S\!(p^*) \circ \revr{\varphi}
      \]
      is rational.
    \end{proof}
  \end{thm}

  \subsection{Ideals}

  A slight generalisation of adding a zero is adding a finite semigroup ideal,
  leading to the following theorem.

  \begin{thm}
    If $S$ is a semigroup such that $S = T \cup I$, and $I$ is a finite ideal,
    then $S$ is a {\rwp-semigroup} if and only if $T$ is a {\rwp-semigroup}.
    \begin{proof}
      Let $S = T \cup I$ be a semigroup such that $I$ is a finite ideal of $S$.

      If $S$ is a {\rwp-semigroup} finitely generated by $\map{p}{A}{S}$, then
      $T$ is finitely generated by $\map{p'}{B}{S}$ where $B = A \cap T$ and $p'$
      is the restriction of $p$ to $B$.
      Theorem \ref{thm:finitely_generated_subsemigroups} implies that $T$ is a
      {\rwp\!-semigroup}.
      
      Conversely, let $T$ be a {\rwp-semigroup} with finite generating system
      $\map{q}{B}{T}$. Let $A = B \cup I$ and $\map{p}{A}{S}$ with $xp = xq$
      if $x \in B$ and $iq = i$ if $x \in I$, hence $p$ is a finite generating
      system for $S$.

      $T$ acts on $I$ on the left via left multiplication, which we denote by
      $\map{\lambda_t}{I}{I}$, and on the right via right multiplication, which
      we denote by $\map{\rho_t}{I}{I}$.
      
      Define the semigroup $U = (\mathcal{T}(I) \times \mathcal{T}(I)) \cup I$
      with multiplication $(\varphi,\psi) \cdot (\varphi',\psi') = (\varphi\varphi',\psi\psi')$,
      $x(\varphi,\psi) = x\varphi$, $(\varphi,\psi)x = x\psi$, and $x\cdot y = xy$.

      The map $\map{r}{A}{U}$ defined by $xr = (\lambda_{xp},\rho_{xp})$ if $x \in B$ and
      $xr = x$ if $x \in I$ uniquely extends to a semigroup morphism $\map{\rho}{A^+}{U}$.

      Using the monoid $U^1 \times U^1$ we see that the relation $\rho_U$ is rational, even
      recognisable. We claim that $\iota_S\!(p) = \iota_T(q) \cup \rho_U$.

      Let $w \in v\iota_S\!(p)$, then $v\pi_p = w\pi_p$ and either $v\pi_p \in T$
      or $v\pi_p \in I$. It follows that either $w \in v\iota_T\!(q)$ or $w \in v\rho_U$, hence
      $w \in v(\iota_S\!(p) \cup \rho_U)$.

      If $w \in v(\iota_S\!(p) \cup \rho_U)$, then $w \in v\iota_S\!(p)$ or $w \in v\rho_U$.
      In the first case $v\pi_q = w\pi_q$ and hence $w \in v\iota_S\!(p)$, in the
      second case $v\rho = w\rho$ and hence $w \in v\rho_U$.
    \end{proof}
  \end{thm}

  \subsection{Direct product}

  In this subsection we characterise when a direct product of finitely generated
  semigroups is an \rwp-semigroup, that is we prove the following theorem.

  \begin{thm}
    \label{thm:direct_product}
    Let $S$ and $T$ be semigroups such that $S \times T$ is finitely generated.
    Then $S \times T$ is an \rwp-semigroup if and only if $S$ and $T$ are
    \rwp-semigroups, and at least one of $S$ or $T$ is finite.
  \end{thm}

  We break up the otherwise slightly unwieldy proof into lemmas. Lemma
  \ref{lem:one_finite} and \ref{lem:two_infinite} for the ``if'' part, and
  Lemma \ref{lem:product_factors} for the ``only-if'' part, and conclude with
  the proof of Theorem \ref{thm:direct_product} at the end of this section.

  First, the direct product of an \rwp-semigroup with a finite semigroup, if
  finitely generated, is a \rwp semigroup.
  \begin{lem}\label{lem:one_finite}
    Let $S$ be a finite semigroup and $T$ be an \rwp-semigroup. If $S \times T$ is
    finitely generated, then $S \times T$ is an \rwp-semigroup.
    \begin{proof}
      Since $S \times T$ is finitely generated, we fix a finite generating system
      $\map{p}{A}{S\times T}$. Then $\map{p\circ\pi_S}{A}{S}$ is a generating system
      for $S$ and $\iota_S\!(p\circ\pi_S)$ is recognisable, and $\map{p\circ\pi_T}{A}{T}$
      is a generating system for $T$ and $\iota_T\!(p\circ\pi_T)$ is rational by assumption.

      It follows that
      \[
      \iota_{S\times T}\!(p) = \iota_S\!(p\circ\pi_S)\cap\iota_T\!(p\circ\pi_T)
      \]
      is a rational relation by Proposition \ref{prop:rec_rat_intersection}.
    \end{proof}
  \end{lem}

  If we take the direct product of two infinite \rwp-semigroups, then even
  if this direct product is finitely generated, it is not an \rwp-semigroup.

  \begin{lem}\label{lem:two_infinite}
    If $S$ and $T$ are infinite \rwp-semigroups such that $S \times T$ is
    finitely generated, then $S \times T$ is not an \rwp-semigroup.
    \begin{proof}
      If $S$ and $T$ are infinite \rwp-semigroups, then by Theorem
      \ref{thm:infinite_order_element} there are elements $s \in S$ and $t \in T$ that
      generate infinite subsemigroups of $S$ and $T$ respectively.

      Consider the subsemigroup of $S \times T$ generated by the two distinct
      elements $x = (s^2,t)$ and $y = (s,t^2)$.

      It holds that $xy = (s^2,t)(s,t^2) = (s^3,t^3) = (s,t^2)(s^2,t) = yx$, hence
      they generate a subsemigroup of $S \times T$ that is isomorphic to
      $\fcs{s,t}$. 

      This contradicts Theorem \ref{thm:finitely_generated_subsemigroups} and
      the fact that $\fcs{s,t}$ is not an \rwp-semigroup.
    \end{proof}
  \end{lem}

  It is a direct consequence of Corollary
  \ref{thm:finitely_generated_subsemigroups} that the factors of a finitely
  generated direct product of \rwp-semigroups are \rwp-semigroups.

  \begin{lem}\label{lem:product_factors}
    If $S$ and $T$ be semigroups such that $S \times T$ is an \rwp-semigroup, then
    $S$ and $T$ are \rwp-semigroups.
    \begin{proof}
      Let $S \times T$ be an \rwp-semigroup and $\map{p}{A}{S\times T}$ a finite
      generating system. If $\map{\pi_S}{S\times T}{S}$ is the projection onto $S$,
      then $p \circ \pi_S$ is a generating system for $S$. The kernel of $p \circ \pi_S$
      is a relation on $A$ which extends to a rational equivalence relation
      $\map{\rho}{\fsg{A}}{\fsg{A}}$. Hence the relation
      \[
      \tau = \rho \circ \iota_{S\times T}\!(p) \circ \rho
      \]
      is rational as a composition of rational relations.

      We show that $\tau = \iota_{S}\!(p \circ \pi_S)$. If $w \in v\tau$,
      then $v\pi_p\pi_S = w\pi_p\pi_S$, hence $w \in v\iota_S\!(p \circ \pi_S)$.

      Conversely, let $w \in v\iota_S\!(p \circ \pi_s)$. This implies that there
      are strings $v'$ and $w'$ such that $w' \in w\rho$ and $v \in v'\rho$, and
      such that $w' \in v'\iota_{S \times T}\!(p)$, hence $w \in v\tau$.
    \end{proof}
  \end{lem}

  For the proof of Theorem \ref{thm:direct_product} let $S$ and $T$ be
  semigroups such that $S \times T$ is finitely generated.
      
  If $S \times T$ is an \rwp-semigroup, then by Lemma \ref{lem:product_factors}
  $S$ and $T$ are \rwp-semigroups.
  Conversely, if both $S$ and $T$ are finite, then $S \times T$ is finite and thus
  an \rwp-semigroup. If $S$ is finite, $T$ is an \rwp-semigroup, and $S \times T$
  is finitely generated, then by Lemma \ref{lem:one_finite} $S \times T$ is a
  \rwp-semigroup.

  If both $S$ and $T$ are infinite, then by Lemma \ref{lem:two_infinite} the
  direct product is not an \rwp-semigroup.

  \subsection{Free product}

  When considering free products, we have to be careful to distinguish which
  category we take the free product in: If we are taking the free product of two
  monoids then there are two categories to consider: the category of semigroups
  and the category of monoids.

  \begin{thm}
    Let $S$ and $T$ be semigroups. Then the free product $S * T$ in the category
    of semigroups is an \rwp-semigroup if and only if $S$ and $T$ are \rwp-semigroups.
    \begin{proof}
      If $S * T$ is an \rwp-semigroup, then $S$ and $T$ are finitely generated
      subsemigroups of $S \star T$, and hence $S$ and $T$ are \rwp-semigroups by
      Theorem \ref{thm:subsemis}.

      Conversely, assume that $S$ and $T$ are \rwp-semigroups, and that $\map{p}{A}{S}$
      and $\map{q}{B}{T}$ are finite generating systems for $S$ and $T$. Then
      $\map{p \cup q}{A \cup B}{S*T}$ is a generating system for $S*T$.
      The relation
      \[
      \rho = \rb{\iota_S(p) \cup \iota_T(q)}^+
      \]
      is rational as a union of rational relations, and equal to
      $\iota_{S * T}(p \cup q)$.
    \end{proof}
  \end{thm}

  We note that in the category of monoids the free product of two \textbf{rwp}-monoids
  is not necessarily a \textbf{rwp}-monoid. For example consider the cyclic group $C_2$
  of order $2$. Then $C_2 * C_2$ is an infinite group, and hence not a \textbf{rwp}-monoid.

  \begin{thm}
    Let $S$ and $T$ be monoids. Then the free product $S * T$ in the category
    of monoids is a \textbf{rwp}-monoid if and only if at least one of $S$ and $T$
    has trivial group of units.
    \begin{proof}
      If $S*T$ is finitely generated and a \rwp-monoid, then $S$ and $T$ are finitely
      generated submonoids of $S*T$, and hence \rwp-monoids.

      If both $S$ and $T$ had non-trivial groups of units, then $S*T$ has an infinite
      subgroup, and hence is not a \rwp-monoid.

      Let without loss of generality $\gu{T}$ be trivial. Then $\gu{S*T} = \gu{S}*\gu{T}$
      is finite, and the set $E = 1\pi_r^{-1}$ is a recognisable subset of $\fm{C}$, and
      the relation $\rela{\rho}{\fm{C}}{\fm{C}}$ which replaces occurrences of elements of
      $E$ by $\varepsilon$ is rational.

      Now the relation
      \[
      \mu = \left(\iota_S\!(p) \cup \iota_T\!(p)\right)^*
      \]
      is rational as a union of rational relations, and the composition $\rho\mu\revr{\rho}$
      is rational as a composition of rational relations.
    \end{proof}
  \end{thm}

  \subsection{Zero union}

  The final construction we consider is the zero union of two semigroups. If $T$
  and $U$ are two semigroups we can form their \emph{zero union} by taking the
  disjoint union $S = T \cup U \cup \{z\}$ of $T$ and $U$ and a zero element $z$.
  Multiplication is defined by
  \[
  x \cdot y = \begin{cases} xy \in T & x \mbox{ and } y \in T \\
    xy \in U & x \mbox{ and } y \in U \\
    z & \mbox{otherwise}.\\
    \end{cases}
  \]
  Conversely, if $S$ is a semigroup such that there exist subsemigroups $T$ and $U$ and an element $z$
  in $S$ such that $S = T \cup U \cup \{z\}$, $T \cap U = \{ \}$, and neither $T$ nor $U$ contain $z$,
  then we say that $S$ is a \emph{zero union} of $T$ and $U$.
  We write $S = T \cup_0 U$ to say that $S$ is a zero union of subsemigroups $T$ and $U$.

  \begin{thm}
    Let $S = T \cup_0 U$ be a zero union of semigroups. Then $S$ is a {\rwp-semigroup} if
    and only if $T$ and $U$ are {\rwp-semigroups}.
    \begin{proof}
      If $S = T \cup_0 U$ is a {\rwp\!-semigroup} then $T$ and $U$ are finitely generated
      subsemigroups of $S$ and hence {\rwp\!-semigroups}.

      Conversely let $T$ and $U$ be {\rwp\!-semigroups} with generating systems $\map{p}{A}{T}$
      and $\map{q}{B}{U}$, and let $S = T \cup_0 U$ be their zero union. Then define a finite
      generating system $\map{r}{C}{S}$ with $C = A \cup B \cup \{z\}$, and
      \[
      r : C \rightarrow S, x \mapsto \begin{cases} xp & x \in A \\ xq & x \in B \\ z & x = z \\ \end{cases}
      \]

      To show that $\iota_S(r)$ is rational we claim that $\iota_S(r) = \iota_T(p) \cup \iota_U(q) \cup \rho$
      where $X = \fsg{A} \cup \fsg{B} \subset \fm{C}$ 
      \[
      \rho = (X \times \{\}) \cup (\{\} \times X) \cup (\overline{X} \times \overline{X}).
      \]
    \end{proof}
  \end{thm}

\section{Decidability}
  \label{sec:decidability}
  
  In this section we consider decision problems for \rwp-semigroups.
 
  Care has to be taken what the input for a decision procedure is.
  Whenever we want to decide a property of an \rwp-semigroup, we assume that the
  word problem is given as a rational expression with respect to a finite
  generating system $\map{p}{A}{S}$, or equivalently a two-tape asynchronous
  finite state automaton.

  While it is undecidable whether a finitely presented semigroup is an \rwp-semigroup,
  once a semigroup is given as an \rwp-semigroup, the word problem is decidable, it is
  decidable whether the semigroup is trivial, a monoid, or a group, and whether
  it is free.
  We will cover further decision problems, such as whether an \rwp-semigroup
  contains a zero, whether it is cancellative, or whether two \rwp-semigroups are
  isomorphic in a forthcoming paper.

  \begin{thm}
    It is recursively undecidable whether a finitely presented semigroup has
    rational word problem.
    \begin{proof}
      Suppose that there exists a Turing machine $M$ that decides whether a given finitely
      presented semigroup has rational word problem.

      Let $S = \spr{A_1}{R_1}$ be a finitely presented monoid with rational word problem and let
      $T = \spr{A_2}{R_2}$ be a finitely presented monoid with undecidable word problem. Note that
      such a monoid exists by \cite{tzeitin2}.
      Let $A = A_1\cup A_2$ and $R = R_1 \cup R_2$. For any $u$ and $v$ in $\fm{A_2}$ define
      \[
      T_{u,v} = \mpr{A,c,d}{R,(cud,\varepsilon),(acvd,cvd) \mbox{ for all } a \in A \cup \{c,d\}}.
      \]

      If $u\pi_{A_2} = w\pi_{A_2}$, then $T_{u,v}$ is trivial, otherwise $T_{u,v}$ has undecidable word
      problem.
      Now the monoid free product $S * T_{u,v}$ has rational word problem if and only if
      $u\pi_{A_2} = v\pi_{A_2}$

      The Turing machine $M$ now decides whether $u\pi_{A_2} = v\pi_{A_2}$, a
      contradiction.
    \end{proof}
  \end{thm}

  \begin{thm}\label{thm:dec finite}
    For a semigroup $S$ given by a rational word problem $\iota_S{p}$, it is
    decidable whether $S$ is finite.
    \begin{proof}
      Let $S$ be a semigroup and $\map{p}{A}{S}$ be a finite generating system.

      Since $\iota_S(p)$ is a rational equivalence relation, by Proposition
      \ref{prop:rational_finite_rep} there is a recognisable subset $D \subset \fm{A}$
      that contains finitely many representatives for each element of $S$.

      $D$ can effectively be obtained from $\iota_S(p)$, and it is decidable
      whether $D$ is finite. \cite{sakarovitch2009}

      The semigroup $S$ is finite if and only if $D$ is, and hence it is
      decidable whether $S$ is finite.
    \end{proof}
  \end{thm}

  It is decidable whether an \rwp-semigroup is a monoid. This proof uses the same
  idea as the proof used in \cite{cainpfeiffer2015}.

  \begin{thm}\label{thm:dec monoid}
    For a semigroup $S$ given by a rational word problem $\iota_S\!(p)$ it is
    decidable whether $S$ is a monoid.
    \begin{proof}
      Let $S$ be a semigroup given by a rational word problem $\iota_S\!(p)$
      with respect to $\map{p}{A}{S}$.

      For each $a \in A$ the languages
      \[
      I_a^r = \{ i \in \fm{A} ~\mid~ ai \in a\iota_{S}\!(p) \}
      \]
      and
      \[
      I_a^l = \{ i \in \fm{A} ~\mid~ ia \in a\iota_{S}\!(p) \}
      \]
      are recognisable subsets of $\fm{A}$, because $\iota_{S}\!(p)$ is a
      rational relation. Hence the intersection
      \[
      I = \bigcap\limits_{a \in A}(I_a^l \cap I_a^r)
      \]
      is recognisable subset of $\fm{A}$.

      Then $S$ is a monoid if and only if $I$ is non-empty.

      If $i \in I$ and if $w = w_1w_2\ldots w_n \in \fm{A}$, then
      \[
      (iw_1w_2\ldots w_n)\pi_p = w\pi_p = (w_1w_2\ldots w_ni)\pi_p,
      \]
      because $(iw_1)\pi_p = w_1\pi_p$ and $(w_ni)\pi_p = w_n\pi_p$.

      Let $S$ be a monoid and let $i \in \fm{A}$ such that $e = i\pi_p$. Then
      $(ai)\pi_p = (ia)\pi_p = a\pi_p$, and hence $I_a^r$ and $I_a^l$ contain $i$,
      so $I$ contains $i$.
    \end{proof}
  \end{thm}

  Finally, it is decidable whether an \rwp-semigroup is a group. This is because we showed that
  an \rwp-semigroup that is also a group has to be finite, and we can decide finiteness and then
  by exhaustive testing find whether the semigroup in question is a group.

  \begin{thm}
    Let $\iota_S(A)$ be a rational word problem. Then it is decidable whether $S$ is a group.
    \begin{proof}
      By Corollary \ref{cor:groupsfinite} for $S$ to be a group it has to necessarily be finite,
      which can be decided by Theorem \ref{thm:dec finite}. Thus, if $S$ is infinite, then it is
      not a group. If it is finite, then by exhaustively checking the group axioms, it can be
      determined whether $S$ is a group.
    \end{proof}
  \end{thm}

  We demonstrate that \rwp-semigroups are word-hyperbolic in the sense of Cain
  and Pfeiffer \cite{cainpfeiffer2015}, in which it is also shown that
  that freeness is decidable for word-hyperbolic semigroups.

  \begin{defn}
    Let $S$ be a semigroup and $\map{p}{A}{S}$ be an injective finite generating
    system. We say that $S$ is \emph{word-hyperbolic} if there exists
    a regular language $L \subset \fsg{A}$, such that
    \[
      M = \left\{ u\#_1v\#_2\#w^{rev} ~\mid~ u,v,w \in L, (u\pi)(v\pi) = w\pi \right\}
    \]
    is context-free.
  \end{defn}

  \begin{thm}\label{thm:wordhyperbolic}
    Every \rwp-semigroup is word-hyperbolic.
    \begin{proof}
      Let $\iota_S(A)$ be a rational word problem. Choose $L = \fsg{A}$, then
      \[
        W = \left\{ v\#w^{rev} ~\mid~ v\pi = w \pi \right\}
      \]
      is context-free.

      We consider the morphism
      \[
        \varphi : \fm{(A \cup \{\#_1,\#_2\})} \rightarrow \fm{(A \cup \{\#\})}
      \]
      defined by
      \[
        x\varphi = 
        \begin{cases}
          x & x \in A\\
          \varepsilon & x = \#_1\\
          \# & x = \#_2
        \end{cases}
      \]

      We consider
      \[
      M = W\varphi^{-1} \cap \fsg{A}\#_1\fsg{A}\#_2\fsg{A},
      \]
      which is context-free because $W\varphi^{-1}$ is the preimage of a
      context-free language under a rational relation, and intersections of
      context-free languages with regular languages are context-free.

      Now $u\#_1v\#_2w^{rev} \in M$ if and only if $(uv)\pi = w\pi$, if and only
      if $(u\pi)(v\pi) =  w\pi$, proving that $S$ is word-hyperbolic.
    \end{proof}
  \end{thm}

  The decision procedures need to take an appropriate input, such as a word-hyperbolic
  structure for $S$, which can be effectively obtained from a rational specification of the
  word problem. The first theorem can be proven directly.

  \begin{cor}
    Let $\iota_S(A)S$ be a rational word problem. Then it is decidable whether $S$ is free.
    \begin{proof}
      By Theorem \ref{thm:wordhyperbolic} $S$ is word-hyperbolic, and a
      word-hyperbolic structure can effectively obtained from $\iota_S\!(A)$, and
      hence by \cite[Section 11]{cainpfeiffer2015} it is decidable whether $S$
      is free.
    \end{proof}
  \end{cor}

  The most important open question in this section is to obtain an undecidable problem for
  a given \rwp-semigroup. The most reasonable candidate to focus on seems to be the isomorphism
  problem.


%
%

\section{Conclusion}

Further work could investigate semigroups with rational Green's relations, or
the structure of Green's relations of \rwp-semigroups more closely. Furthermore
intersections of rational relations should be more closely investigated, as for
instance the word problem of $\fcs{a,b}$ is the intersection of two rational
relations. Furthermore for an \rwp-semigroup, the Green's relations
are intersections of two or more rational relations.

The advanced results in \cite{DBLP:journals/iandc/Sakarovitch87,DBLP:journals/iandc/PelletierS90}
need to be put into the context of our definition of \rwp-semigroups.

The question whether \rwp-semigroups coincide with rational semigroups,
equivalently whether rational equivalence relations have regular cross sections
needs to be settled.

As an application of the methods presented in this paper, an implementation of
algorithms using finite state automata to compute with \rwp-semigroups would be
desirable.

The second author would like to thank Jacques Sakarovitch for his very helpful
comments and the coining of the term \rwps.

%
  
\bibliographystyle{model1-num-names} 
\bibliography{word_problems_fsa_2}

\end{document}